\documentclass[reqno,10pt,centertags,draft]{amsart}
\usepackage{amsmath,amsthm,amscd,amssymb,latexsym,upref}

\def\alg{{\mathfrak A}}

\def\cref{c_{ref}}
\def\cuts{\kappa_\sig}

\def\e{\varepsilon}

\def\Fo{{\mathfrak F}}

\def\g{{\sqrt\alpha}}
\def\gs{\alpha}

\def\H{{\mathcal H}}

\def\nvec{u}

\def\Psivec{\Psi_{\nvec}}
\def\puppbd{{\frac{1}{3}}}
\def\puppbdpi{{\frac{1}{20}}}

\def\sig{{\sigma}}

\def\vac{\Omega_f}

\def\bra{\big\langle}
\def\ket{\big\rangle}
\def\Bra{\Big\langle}
\def\Ket{\Big\rangle}

\def\C{{\Bbb C}}
\def\N{{\Bbb N}}

\def\R{{\Bbb R}}

\def\cB{{\mathcal B}}

\def\1{{\bf 1}}

\def\eqnn{\begin{eqnarray*}}
\def\eeqnn{\end{eqnarray*}}
\def\eqn{\begin{eqnarray}}
\def\eeqn{\end{eqnarray}}

\def\bal{\begin{align}}
\def\eal{\end{align}}

\theoremstyle{plain}
\newtheorem{theorem}{Theorem}[section]

\newtheorem{proposition}[theorem]{Proposition}

\newtheorem{lemma}[theorem]{Lemma}

\numberwithin{equation}{section}

\def\prf{\begin{proof}}
\def\endprf{\end{proof}}

\begin{document}

\bibliographystyle{plain}

\title[Infrared representations in non-relativistic QED]
{Coherent infrared representations in non-relativistic QED}

\author[T. Chen]{Thomas Chen}
\address{Department of Mathematics,
Princeton University, 807 Fine Hall, Washington Road, Princeton,
NJ 08544, U.S.A.}
\email{tc@math.princeton.edu}

\author[J. Fr\"ohlich]{J\"urg Fr\"ohlich}
\address{Institute for Theoretical Physics,
ETH H\"onggerberg, 8093 Z\"urich, Switzerland.}
\email{juerg@itp.phys.ethz.ch}


\begin{abstract}
We consider dressed 1-electron states in a translation-invariant model
of non-relativistic QED.
To start with a well-defined model, the interaction Hamiltonian
is cutoff at very large photon energies (ultraviolet cutoff) and
regularized at very small photon energies (infrared regularization). The
infrared regularization is then removed, and the
representations of the canonical commutation relations
of the electromagnetic field operators determined by the dressed
1-electron states are studied using operator-algebra methods.
A key ingredient in our analysis is a bound on the renormalized
electron mass  uniform in the infrared regularization.
Our results have important applications in the scattering theory for
infraparticles.
\end{abstract}

\maketitle

\parskip = 8 pt

Dedicated to Barry Simon on the occasion of his 60th birthday,
in admiration and friendship.

{\renewcommand{\baselinestretch}{.1} \scriptsize \tableofcontents}

\section{Introduction}
\label{sec:intro}

In this note, we consider a translation-invariant model of non-relativistic
Quantum Electrodynamics (QED) describing a non-relativistic
Pauli (spin $\frac12$) electron
interacting with the quantized electromagnetic field.
An infrared regularization (parametrized by a number $\sigma\ll1$)
and  a fixed ultraviolet cutoff are imposed on the interaction Hamiltonian.
Let $H(p,\sigma)$ denote the cutoff fiber Hamiltonian
corresponding to the conserved momentum $p$ on the fiber Hilbert space $\H_p$.
This space is isomorphic to $\C^2\otimes\Fo$, where $\Fo$ denotes the photon Fock space,
and $\C^2$ accounts for the spin of the electron. It is
proved in \cite{bcfs2} and \cite{ch1} that, for sufficiently small values
of the finestructure constant, $H(p,\sigma)$ possesses a  ground state
eigenvalue $E(p,\sigma)$ (of multiplicity two for spin $\frac12$)
at the bottom of its essential spectrum. Let ${\mathcal E}_{p,\sigma}$ denote the 
corresponding ground state eigenspace.
The unit rays determined by the eigenvectors  $\Psivec(p,\sigma)\in{\mathcal E}_{p,\sigma}$,
$\|\Psivec(p,\sigma)\|=1$, can be parametrized
by $u\in S^2\subset\R^3$, with
$\bra \Psivec(p,\sigma) \, , \, \tau \, \Psivec(p,\sigma) \ket = \nvec$
($\tau$ is the vector of Pauli matrices, see (\ref{eq:pauli-def-1})).

Let $K_\rho:=\{k\in\R^3\,\big|\,|k|\geq\rho\}$ be the set of photon momenta corresponding
to photon energies $\geq\rho$.
(We choose units such that $\hbar=c=1$. The finestructure constant is $\gs=e^2$.)
By $\Fo_\rho$ we denote the symmetric Fock space over
the one-photon Hilbert space $L^2(K_\rho, d^3k)\otimes \C^2$ of
wave functions describing the pure states of a photon of
energy $\geq\rho$; the factor $\C^2$ accounts for the
two possible polarizations of a photon.
Let $\cB(\Fo_\rho)$ denote the algebra of all bounded operators on $\Fo_\rho$. We define
a $C^*$-algebra, $\alg$, by setting
$$
        \alg:=\overline{\bigvee_{\rho>0}\cB(\Fo_\rho)}^{\|\,\cdot\,\|} \;,
$$
where the closure is taken in the operator norm.
We are interested in the representations of $\alg $ determined by dressed
1-electron states via the GNS construction. We define the infrared-regularized states
$$
        \omega_{p,\sigma}(A) \; := \; \langle \, \Psivec(p,\sigma) \, , \, A \, \Psivec(p,\sigma) \, \ket
        \;\;\;,\;\;\;A\in\alg \;,
$$
for a fixed choice of $u\in S^2$.
We prove that, for momenta $p$ with $0\leq |p|<\puppbd$ and any sequence $\sigma_n\searrow0$
($n\rightarrow\infty$), there exists a state $\omega_p$ on
$\alg$ given by $\omega_p(A)=\lim_{j\rightarrow\infty}\omega_{p,\sigma_{n_j}}(A)$,  for all
$A\in\alg$, for some subsequence $(\sigma_{n_j})$.
By the GNS construction, the state $\omega_p$ determines a  representation of $\alg$.
For $p\neq0$, this representation turns out to be
quasi-equivalent to a {\em coherent state representation} of $\alg$
{\em unitarily inequivalent} to the Fock representation.
It will be determined explicitly.

For Nelson's model, similar results were proven in \cite{fr,fr2}.
However, the more complicated coupling structure of the Hamilton operator of
non-relativistic QED makes a key argument in \cite{fr} inapplicable.
The difficulty arises from the fact that the interaction
term in QED is of minimal substitution type
and hence {\em quadratic} in creation- and annihilation operators,
while, in Nelson's model, it is linear.
We arrive at our main result by making use of
the {\em uniform bounds on the renormalized electron mass}
recently derived in \cite{ch1} and \cite{bcfs2}.

An important application of our results concerns infraparticle scattering theory,
in particular Compton scattering.
Recently, some significant progress in scattering theory
was made by A. Pizzo in \cite{pi2},
where infraparticle scattering states are constructed
for Nelson's model after a complete removal of the infrared regularization.
The proof uses, and significantly extends, ideas proposed in \cite{fr,fr2}.
A bound on the renormalized particle mass uniform in the infrared
regularization $\sigma\geq0$ is assumed
in \cite{pi2} without proof.

The construction of an infraparticle scattering state in \cite{fr,pi2} crucially
involves a {\em dressing transformation}.
To construct the latter, it is necessary to identify a coherent state representation
that is quasi-equivalent to the GNS representation determined by $\omega_p$.
This was achieved in \cite{fr} for Nelson's model,
but has not been accomplished for non-relativistic QED,
due to the difficulties noted above.
This is the main reason why attempts to construct an infraparticle
scattering theory for non-relativistic QED have been unsuccessful, so
far, even after the appearance of Pizzo's work.
With Theorem {\ref{thm:main-2}} of the present paper, we provide this important
missing ingredient. Further modifications necessary to adapt Pizzo's analysis to non-relativistic
QED are outlined, but
a detailed discussion of these matters is beyond the scope of the present paper.

\section{Definition of the model}
\label{sec:model-def}

We consider an electron of spin $\frac12$ coupled to the quantized electromagnetic field,
with a fixed ultraviolet cutoff imposed on the interaction Hamiltonian.

The  Hilbert space of one-electron states is given by
\eqn
    \H_{el} \; = \; L^2(\R^3)\otimes \C^2   \;.
\eeqn
The Fock space of the quantized electromagnetic field in the Coulomb gauge is given by
\eqn
    \Fo \; = \; \bigoplus_{n\geq0}\Fo^{(n)}
    \;\;\;,\;\;\;
    \Fo^{(0)} \; = \; \C\;,
\eeqn
where the fully symmetrized $n$-fold tensor product space
\eqn
    \Fo^{(n)} \; = \; {\rm Sym}_n(L^2(\R^3) \otimes \C^2)^{\otimes n}
\eeqn
denotes the $n$-photon Hilbert space.
The factor $\C^2$ accounts for the
two transverse polarization modes of a photon, and ${\rm Sym}_n$
symmetrizes the $n$ factors in the tensor product,
in accordance with the fact that photons are bosons.

A vector $\Phi\in\Fo$ corresponds to a sequence
$$
        \Phi \; = \; (\Phi^{(0)},\Phi^{(1)},\dots,\Phi^{(n)},\dots) \; \; , \; \;
        \Phi^{(n)}\in\Fo^{(n)}  \;,
$$
where $\Phi^{(n)}=\Phi^{(n)}(k_1,\lambda_1,\dots,k_n,\lambda_n)$,
$k_j\in\R^3$ is the momentum, and $\lambda_j\in\{+,-\}$
labels the two possible helicities of the $j$-th photon.
The scalar product on $\Fo$ is given by
\eqnn
        \bra \, \Phi_1 \, , \, \Phi_2 \, \ket \; = \; \sum_{n\geq0}
        \bra \, \Phi_1^{(n)} \, , \,
        \Phi_2^{(n)} \, \ket_{\Fo^{(n)} } \;.
\eeqnn
Let $\widehat f$ denote the Fourier transform of $f$.
For $\lambda\in\{+,-\}$ and  $f\in L^2(\R^3)$,
we introduce annihilation operators
\eqn
        a_\lambda(f) \, : \, \Fo^{(n)} \; \rightarrow \; \Fo^{(n-1)}
\eeqn
defined by
\eqn
        &&(a_\lambda(f) \, \Phi)^{(n-1)}(k_1,\lambda_1,\dots,k_{n-1},\lambda_{n-1})
        \nonumber\\
        &&\hspace{1cm}= \sqrt n \,
        \int d k_n \, \widehat f^*(k_n) \, \Phi^{(n)}(k_1,\lambda_1,\dots,k_{n-1},
        \lambda_{n-1},k_n,\lambda )
        \label{eq:ann-op-def-1}
\eeqn
and creation operators
\eqn
        a_\lambda^*(f) \, : \, \Fo^{(n)} \; \rightarrow \; \Fo^{(n+1)}
        \;\;\;,\;\;\;{\rm with}\;\;
        a_\lambda^*(f)= (a_\lambda(f))^*\;.
\eeqn
These operators satisfy the canonical commutation relations
\eqn
      \big[a_\lambda(f),a_{\lambda'}^*(g )\big] &=&
      ( f , g )_{L^2} \delta_{\lambda, \lambda'}
        \;
      \nonumber\\
      &&
      \nonumber\\
      \big[a^\sharp_\lambda (f ),a^\sharp_{\lambda'} (g )\big] &=& 0 \;,
\eeqn
for all $f,g\in L^2(\R^3)$, where $a^\sharp$ denotes either $a_\lambda$ or $a_\lambda^*$.
The {\em Fock vacuum} is the unique unit vector
\eqn
        \vac \; = \; (1,0,0,\dots)
\eeqn
in $\Fo$ with the property that
\eqn
        a_\lambda(f) \, \vac \; = \; 0 \;,
\eeqn
for all $f\in L^2(\R^3)$.

Since $a^*_\lambda(f)$ is linear and $a_\lambda(f)$ is antilinear in $f$, one can write
\eqn
        a_\lambda^*(f)
        \; = \; \int_{\R^3}d k \,  a_\lambda^*(k ) \widehat f(k)
        \; \; , \; \;
        a_\lambda(f)
        \; = \; \int_{\R^3}d k \, \widehat f^*(k) a_\lambda(k )
\eeqn
where
$a_\lambda^\sharp(k)$ are operator-valued distributions
also referred to as creation- and annihilation operators.
They satisfy the commutation relations
\eqn
        \big[a_{\lambda'}( k' ),a^*_\lambda( k )\big] &=& \delta_{\lambda, \lambda'} \,
        \delta ( k- k')
        \nonumber\\
        &&
        \nonumber\\
        \big[a^\sharp_{\lambda'} ( k' ),a^\sharp_\lambda ( k )\big] &=& 0
\eeqn
for all $ k,  k'\in\R^3$ and $\lambda,\lambda'\in\{+,-\}$, and
\eqn
        a_\lambda( k )\,\vac=0
\eeqn
for all $ k$, $\lambda$.

The Hilbert space of the system consisting of a single Pauli electron
and the quantized radiation field is given by the tensor product space
\eqn
    \H \; = \; \H_{el}\otimes\Fo \;.
\eeqn
The Hamiltonian is given by
\begin{align}
    H(\sigma) \; = \; \frac12 \, \big(i\nabla_x\otimes\1_f \, - \, \g \,  A_\sigma(x)
    \big)^2 \, + \, \g \, \tau\cdot B_\sigma(x)
    \, + \, \1_{el}\otimes H_f \;,
\end{align}
where
\eqn
    H_f&=& \sum_{\lambda=1,2}\int dk \, |k| \, a_\lambda^*(k) \, a_\lambda(k)
\eeqn
is the free-field Hamiltonian, and
\begin{align}
    A_\sigma(x)&= \; \sum_{\lambda=+,-}\int\frac{dk}{\sqrt{|k|}}
    \, \cuts(|k|) \,
    \Big[\epsilon_\lambda(k) e^{-ikx}\otimes a_\lambda(k)+h.c.\Big]
    \nonumber\\
    B_\sigma(x)&= \; \sum_{\lambda=+,-}\int\frac{dk}{\sqrt{|k|}}
    \, \cuts(|k|) \,
    \Big[ (-ik)\wedge\epsilon_\lambda(k) e^{-ikx}\otimes a_\lambda(k)+h.c.\Big] \;.
    \label{eq:A-B-def-1}
\end{align}
denote the (ultraviolet-cutoff) quantized electromagnetic vector potential in the Coulomb gauge,
and the magnetic field operator, respectively.
The function $\cuts$ imposes an ultraviolet cutoff and
an infrared regularization parametrized by $\sigma\ll1$. One may choose it to satisfy
\eqn
    \;\;\;\;\;\;\;
    \cuts(x)=\left\{
    \begin{array}{rl}
    x/\sigma &{\rm for}\; \; x\in[0,\sigma]
    \\
    1 & {\rm for}\; \;  x\in[ \sigma,\frac12],\;  C^\infty \;
    {\rm and \; positive, \; for} \; \;
    x\in (\frac12,1),
    \\
    0 & {\rm for}\; \; x>1 \;.
    \end{array}
    \right.
    \label{eq:cuts-def-1}
\eeqn
Moreover, $\tau=(\tau_1,\tau_2,\tau_3)$, with
\eqn
    \tau_1
    =\left(
    \begin{array}{cc}0&1\\1&0\end{array}\right) \; \; , \; \;
    \tau_2
    =\left(
    \begin{array}{cc}0&i\\-i&0\end{array}\right) \; \; , \; \;
    \tau_3
    =\left(
    \begin{array}{cc}1&0\\0&-1\end{array}\right) \; \; ,
    \label{eq:pauli-def-1}
\eeqn
denotes the vector of Pauli matrices.
We remark that $H(\sigma)$ defines a selfadjoint operator on $\H$ bounded
from below; see e.g. \cite{resi2}.

For a charged particle of spin $0$, the Zeeman term proportional to $B_\sigma(x)$ is absent.

The momentum operator of the system is given by
\eqn
    P_{tot} \; = \; i\nabla_x\otimes\1_f \, + \, \1_{el}\otimes P_f \;,
\eeqn
where
\eqn
    P_f \; = \; \sum_{\lambda=+,-} \, \int dk \, k \, a_\lambda^*(k) \, a_\lambda(k)
\eeqn
is the momentum operator of the electromagnetic field.

The model under consideration is translation invariant in the sense that
\eqn
        [H(\sigma),P_{tot}] \; = \; 0 \;.
\eeqn
We decompose the Hilbert space
into a direct integral,
\eqn
    \H \; = \; \int_{\R^3}^\oplus dp \, \H_p \;,
\eeqn
where  $\H_p$, the fiber Hilbert space corresponding to a total momentum $p$, is
isomorphic to $\C^2\otimes \Fo$.
Since $\H_p$ is invariant under $\exp[-itH(\sigma)]$, we may consider the
restriction of $H(\sigma)$ to $\H_p$,
\begin{align}
    H(p,\sigma) \; = \; H(\sigma)\Big|_{\H_p}
    \; = \; \frac12 \, \big(p \, - \, P_f \, - \, \g A_\sigma\big)^2 \,
    + \, \g \, \tau\cdot B_\sigma \, + \, H_f \;,
\end{align}
where, henceforth, $A_\sigma\equiv A_\sigma(0)$ and $B_\sigma\equiv B_\sigma(0)$.

We will use results established in \cite{ch1} and \cite{bcfs2} on the nature of the
infimum of the spectrum of $H(p,\sigma)$, for $|p|$ sufficiently small. We define
\eqn
    E(p,\sigma) \; := \; {\rm infspec}\{ \, H(p,\sigma) \, \} \;.
\eeqn
The following theorem is proved in \cite{ch1,ch2}.

\begin{theorem}
\label{thm:main-recall}
Assume that $|p|<\puppbd$. There exists a  small positive constant $\gs_0$
independent of $\sigma$ such that,
for all $\gs<\gs_0$, the following holds: For every $\sigma>0$, $E(p,\sigma)$ is an
eigenvalue at the bottom of the essential spectrum, and, by rotation symmetry,
is a function only of $|p|$. The corresponding eigenspace ${\mathcal E}_{p,\sigma}$
has dimension $2$ for spin $\frac12$.

The functions $E(p,\sigma)$, $\partial_{|p|}E(p,\sigma)$ and $\partial_{|p|}^2E(p,\sigma)$  are
uniformly bounded in $\sigma\geq0$.
There is a constant $c_0>0$ independent of $\sigma$ and $\gs$ such that
the second derivative
\begin{align}
    \partial_{|p|}^2 E(p,\sigma) &= & 1 \, -\, 2\,\bra \, \nabla_p\Psivec(p,\sigma)\,,\,
    (H(p,\sigma)-E(p,\sigma)) \, \nabla_p\Psivec(p,\sigma) \, \ket \;,
\end{align}
where $\Psivec(p,\sigma)\in {\mathcal E}_{p,\sigma}$, $\|\Psivec(p,\sigma)\|=1$,
$\nvec\in S^2$ (see Section {\ref{sec:intro}}),
satisfies
\eqn
    1 \, - \, c_0\gs \; < \; \partial_{|p|}^2 E(p,\sigma) \; < \; 1 \;,
\eeqn
and
\eqn
    \big| \, E(p,\sigma) \, - \, \frac{p^2}{2} \, - \,
    \frac{\gs}{2} \, \bra \, \vac \, , \, A_\sigma^2 \, \vac \, \ket\, \big|
    & < &
    \frac{c_0 \, \gs \, p^2}{2} \;,
    \nonumber\\
    \big| \, \nabla_p E(p,\sigma) \, - \, p \, \big| &<&c_0 \, \gs \, |p| \;.
    \label{eq:derpE-bd-1}
\eeqn
\item
The renormalized electron mass,
\eqn
    m_{ren}(p,\sigma) \; := \; \frac{1}{\partial_{|p|}^2E(p,\sigma)} \;,
\eeqn
is bounded by
\eqn
    1 \; < \; m_{ren}(p,\sigma) \; < \; 1 \, + \, c_0  \gs \;,
    \label{eq:renmass-bd}
\eeqn
uniformly in $\sigma\geq0$, i.e., the radiative corrections {\em increase}
the mass of the electron by an amount of $O(\gs)$.
\end{theorem}

In \cite{bcfs2}, a convergent,  finite algorithm is devised to determine
$m_{ren}(0,0)$ to any given precision, with rigorous error bounds.

We remark that (\ref{eq:derpE-bd-1}) implies that $\nabla_p E(p,\sigma)=0$
if and only if $p=0$, for all momenta $p$, with $|p|<\puppbd$,  all $\sigma\geq0$,
and $\gs$ sufficiently small.

\section{Statement of the main Theorems}

In this paper, we prove accurate upper and lower bounds on the expected photon number
in the dressed one-electron state $\Psivec(p,\sigma)$
and study the GNS representation determined by $\Psivec(p,\sigma)$
in the limit $\sigma\searrow0$, for momenta $p$ with $|p|<\puppbd$.

\subsection{Estimates on the expected photon number}

Our first main result is the following theorem:

\begin{theorem}
\label{thm:main-1}
Assume that $|p|<\puppbd$, and let
$$
        N_f \; = \; \sum_\lambda\int dk \; a_\lambda^*(k) \, a_\lambda(k)
$$
denote the photon number operator.
Then, for all $\gs<\gs_0$ (where $\gs_0$ is the same constant as in Theorem {\ref{thm:main-recall}}),
and independently of $u\in S^2$,
the following holds.

For $p\neq0$, so that $\nabla_p E(p,\sigma)\neq0$,
\eqn
    \Big(-c\gs + c'\gs|\nabla_pE(p,\sigma)|^2\log\frac1\sigma\Big)_+
    & \leq &
    \bra \, \Psivec(p,\sigma) \, , \, N_f \, \Psivec(p,\sigma) \, \ket
    \nonumber\\
    & \leq &
    C\gs + C'\gs|\nabla_pE(p,\sigma)|^2\log\frac1\sigma
    \label{eq:Nf-divergence-sharp}
\eeqn
for non-negative constants $c$, $C$, and $0<c' < C'$ independent of $p$, $\gs$ and $\sigma\geq0$;
(here $r_+:=\max\{0,r\}$).

For $p=0$ (with $\nabla_p E(0,\sigma)=0$),
\eqn
    \bra \, \Psivec(0,\sigma) \, , \, N_f \, \Psivec(0,\sigma) \, \ket
    \; \leq \;
    C\,\gs \;,
\eeqn
uniformly in $\sigma\geq0$.
\end{theorem}

\subsection{Infrared representations}

For $\rho>0$, let
\eqn
    \alg_{\rho } \; := \; \cB(\Fo_\rho)
\eeqn
denote the algebra of bounded operators on
\eqn
    \Fo_\rho \; := \; \bigoplus_{n\geq0}\Fo_\rho^{(n)}
\eeqn
with
\eqn
    \Fo_\rho^{(n)} \; := \; {\rm Sym}_n(L^2(K_\rho,dk)\otimes\C^2)^{\otimes n} \;,
\eeqn
where
\eqn
    K_\rho \; := \; \big \{ \, k\in\R^3\,\big|\,|k|\geq\rho \,\big \} \;.
\eeqn
As indicated above, we define a $C^*$ algebra $\alg$ as the direct limit
\eqn
    \alg  \; := \; \overline{\bigvee_{ \rho > 0 }\alg_{\rho }}^{\;\|\,\cdot\,\|} \;,
\eeqn
where $\overline{(\;\cdot\;)}^{\;\|\,\cdot\,\|}$ denotes the closure with respect to the
operator norm.

We define a state $\omega_{p,\sigma}$ on $\alg$ by
\eqn
    \omega_{p,\sigma}(A) \; = \; \bra\Psivec(p,\sigma) \, , \, A\Psivec(p,\sigma)\ket \;,
\eeqn
for $A\in\alg$,
corresponding to a vector $\Psivec(p,\sigma)\in {\mathcal E}_{p,\sigma}$ (the
space of dressed 1-electron states, i.e., ground states of the fiber
Hamiltonian $H(p,\sigma)$). The choice of $u\in S^2$ is arbitrary but fixed;
our results will not depend on $u$.

We prove that, in the limit $\sigma\searrow0$, $\omega_{p,\sigma}$ tends to a well-defined
state, $\omega_p$, on $\alg$ which determines a GNS representation that is quasi-equivalent to a
{\em coherent state representation}.

\begin{theorem}
\label{thm:main-2}

Assume that $0\leq|p|<\puppbd$, and let $\omega_{p,\sigma}$ be as defined above.
Then, for all $\gs<\gs_0$ (where $\gs_0$ is the same constant as in Theorem {\ref{thm:main-recall}}),
the following holds.

\begin{itemize}

\item[1.]
Let $\{\sigma_i\}$ denote an arbitrary sequence with
$\lim_{i\rightarrow\infty}\sigma_i=0$.
Then there exists a subsequence $\{\sigma_{i_j}\}$ and
a state $\omega_p$ on $\alg$ such that
\eqn
    \lim_{j\rightarrow\infty}\omega_{p,\sigma_{i_j}}(A)=\omega_p(A) \;,
\eeqn
for all $A\in\alg$.
The state $\omega_p$ is normal on the subalgebras $\alg_\rho$, for $\rho>0$.
\\

\item[2.]
The state $\omega_{p,\sigma}$ satisfies
\begin{align}
    \Big|\,\omega_{p,\sigma }( a_\lambda(k)^*  a_\lambda(k))
    \, - \, |\omega_{p,\sigma }(a_\lambda(k))|^2\,\Big|
    \; \leq \; c \, \gs \, \frac{\cuts^2(|k|)}{|k|^{\frac52}}
    \label{eq:state-int-1}
\end{align}
uniformly in $\sigma\geq0$, where $\cuts$ is the cutoff function (\ref{eq:cuts-def-1}),
and
\eqn
    \int dk\,\Big|\,\omega_{p }( a_\lambda(k)^*  a_\lambda(k))
    \, - \, |\omega_{p }(a_\lambda(k))|^2\,\Big|
    \; \leq \; C \, \gs \;,
\eeqn
in the limit $\sigma\searrow0$, for some finite constants $c$, $C$.
\\

\item[3.]
Let $\pi_p$ denote the representation of $\alg$,
$\H_{\omega_p}$ the Hilbert space, and $\Omega_p\in\H_{\omega_p}$
the cyclic vector corresponding to $(\omega_p,\alg)$ by the GNS construction,
(with $\omega_p(A)=\bra\Omega_p\,,\,\pi_p(A)\Omega_p\ket$, for all $A\in \alg$).
Moreover, let
\begin{align}
    v_{p,\sigma,\lambda}(k ) \; := \; - \, \g \, \e_\lambda(k)\cdot\nabla_p E(p,\sigma) \,
    \frac{\cuts(|k|)}{|k|^{\frac12}}\frac{1}{|k|-k\cdot\nabla_p E(p,\sigma)} \;,
    \label{eq:vpkernel-def}
\end{align}
and
\eqn
    v_{p,\lambda}(k ) \; := \; \lim_{\sigma\searrow0}v_{p,\sigma,\lambda} (k) \;.
\eeqn
Then,  $\pi_p$ is quasi-equivalent to $\pi_{Fock}\circ\alpha_p$ (where
$\pi_{Fock}$ is the Fock representation of $\alg$), and
$\alpha_p$ is the *-automorphism of $\alg$
determined by
\eqn
    \alpha_p(a_\lambda^\sharp(k)) \; = \;
    a_\lambda^\sharp(k) \, + \, v_{p,\lambda}^\sharp(k) \;.
    \label{eq:coh-state-1}
\eeqn

\item[4.]
The  Fock representation and $\pi_p$ are related to each other as follows.

(i) If $p=0$
\eqn
    |\lim_{\sigma\searrow0} \omega_{0,\sigma}(N_f)| \; < \; c \, \gs \;,
    \label{eq:num-lim-2}
\eeqn
and $\pi_0$ is (quasi-)equivalent to $\pi_{Fock}$.

(ii) If $p\neq0$, $\pi_p$ is unitarily inequivalent to the Fock representation, and
\eqn
    \lim_{\sigma\searrow0} \omega_{p,\sigma}(N_f) \; = \; \infty \;.
    \label{eq:num-lim-1}
\eeqn
However, $\omega_p$ has the following "local Fock property":
\begin{itemize}
\item[(a)]
For every $\rho>0$,
the restriction of $\omega_p$ to $\alg_\rho$ determines a GNS representation which
is quasi-equivalent to the Fock representation.
\item[(b)]
For every bounded
region $B$ in physical $x$-space, the restriction of $\omega_p$ to the local algebra
$\alg(B)$ determines a GNS representation which is quasi-equivalent to the
Fock representation of $\alg(B)$.
\end{itemize}

\end{itemize}
\end{theorem}

Similar results also hold for a charged particle with spin 0.

\section{Infraparticle scattering}

In this section, we comment on the significance and implications of our results for
the scattering theory of infraparticles, more precisely Compton scattering, in view of
recent work of A. Pizzo.

A framework for an infraparticle scattering theory
in Nelson's model was outlined in \cite{fr},
and the existence of one-electron scattering states for $\sigma>0$ was established.
The existence of scattering states in the
limit $\sigma\searrow0$ has only recently been proven by Pizzo
for Nelson's model in \cite{pi2}, using results in \cite{pi}.

The only unproven hypothesis in \cite{pi2} is that the renormalized electron mass
satisfies $m_{ren}(p,\sigma)<c$, {\em uniformly} in $\sigma>0$,
for $|p|<\puppbdpi$ (in our units).
Uniform bounds on the renormalized electron mass in non-relativistic QED
are proven in  \cite{ch1,ch2} and \cite{bcfs2}, and also hold for Nelson's
model;  (but it has to be assumed there that the infrared regularization
$\cuts(|k|)$ is non-zero in an open neighborhood of $|k|=0$).
The infrared regularization in \cite{pi2} is implemented by a sharp
cutoff $\chi(|k|>\sigma)$, because \cite{pi2} uses results of \cite{fr,pi}
(where this choice is technically convenient).
Replacing  $\chi(|k|>\sigma)$ by  $\cuts(|k|)$
in \cite{pi2} can be implemented with minor modifications.
The methods of \cite{ch1,ch2} then yield the bound $\partial_{|p|}^2E(p,\sigma)>c$,
uniformly in $\sigma\geq0$.

An inequality similar to (\ref{eq:state-int-1}) for Nelson's model plays a central r\^ole
in \cite{fr} and \cite{pi2}, since it explicitly identifies a coherent state representation
which is quasi-equivalent to the GNS representation defined by $\omega_p$.
This coherent state representation determines the correct choice of a {\em "dressing transformation"}
for the asymptotic (free) comparison dynamics,
which is an essential ingredient for the construction of infraparticle scattering states.
With (\ref{eq:state-int-1}), we provide such a dressing transformation for non-relativistic QED.
However, due to the more complicated
structure of the interaction Hamiltonian in non-relativistic QED, as compared to Nelson's model,
there are some additional modifications which
we sketch without detailed proofs.

We start by recalling some basic results in \cite{pi2}, but formulated for QED. Let
\eqn
    \Sigma\; := \;
    \Big\{p\in\R^3\,\Big|\,|p|<\puppbd\Big\}
\eeqn
denote the ball of admissible infraparticle momenta (in \cite{pi2}, the bound $|p|<\puppbdpi$ is used).
Let
\begin{align}
    W_{p,\sigma}(t) \; := \; e^{-itH_f}
    \exp\Big[ i\sum_{\lambda=+,-}\Pi_\lambda( v_{p,\sigma,\lambda})\Big] e^{itH_f} \;,
    \label{eq:Wpsig-dress-1}
\end{align}
where $t$ denotes time. The function $v_{p,\sigma,\lambda}(k)$ is defined in (\ref{eq:vpkernel-def}), and
$\Pi_\lambda(f):=i(a_\lambda(f)-a_\lambda^*(f))$.
As proposed in \cite{fr},
a natural candidate for the asymptotic, freely moving comparison state is given by
\begin{align}
    &&\int_{\Sigma} dp \; W_{p,\sigma }(t) \, e^{-i(p-P_f)x} \,
    e^{i\gamma_{\sigma }(\nabla_pE(p,\sigma ),t)}
    \,
    h(p)  \, e^{-itE(p,\sigma )} \, \Psivec(p,\sigma )
    \;\;\in\,\H_{el}\otimes\Fo\;.
    \label{eq:freestate-def}
\end{align}
Here, $h(p) \, e^{-itE(p,\sigma)} \, \Psivec(p,\sigma)$ describes a freely
moving electron with wave function $h$ (in Schwartz space, and supported in $\Sigma$).
The operator $W_{p,\sigma}(t)$ describes a freely time-evolving
cloud of physical soft photons surrounding the electron.
The integral over $p$ and the factor $e^{-i(p-P_f)x}$ implement the inverse Fourier
transform. The purpose of adding a scalar phase factor $\gamma_{\sigma }(\nabla_pE(p,\sigma ),t)$
(which we do not specify in detail here)
is similar as in Dollard's classical construction of modified wave
operators for Coulomb scattering, \cite{do}.
While the limit $\sigma\searrow0$ of the one-electron states $\Psivec(p,\sigma)$ does
not define vectors in the Fock spaces $\H_p$, the limit $\sigma\searrow0$ of
the vectors (\ref{eq:freestate-def}) defines vectors in the physical Hilbert space $\H_{el}\otimes\Fo$.

Next, we sketch the main construction in \cite{pi2}.

In \cite{pi2}, a discretized (Riemann sum) version of (\ref{eq:freestate-def}) is used
as the free comparison state, where the resolution of the discretization becomes
arbitrarily fine, as $t\rightarrow\infty$.

Let $T_n^{(\e)}:=2^{n/\e}$, for some $0<\e\ll1$ (to be fixed appropriately). Let
\eqn
    \Sigma \; = \;  \bigcup_{j=1}^{N(t)}\Gamma_j(t)
\eeqn
be a decomposition of $\Sigma$ into
disjoint cubic cells, where the number of cells $N(t)$ is time-dependent,
and given by
\eqn
    N(t) \; = \; (2^n)^3 \; \; , \; \; {\rm for } \; \; \; \;
    T_n^{(\e)} \; \leq \; t \; < \; T_{n+1}^{(\e)} \;.
    \label{eq:Nt-def}
\eeqn
The analysis in \cite{pi2} assumes that
\eqn
    \partial_{p}^2E(p,\sigma)=(m_{ren}(p,\sigma))^{-1} \; > \; c \; > \; 0 \;,
\eeqn
uniformly in $\sigma\geq0$.

A key element of the construction in \cite{pi2} is to render the infrared cutoff $\sigma_t$ {\em time-dependent},
with $\sigma_t$ converging to $0$ at a prescribed rate, as $t\rightarrow\infty$.

Accordingly, let
\eqn
    \psi_{h,\sigma_t}(t,x) \; := \; \sum_{j=1}^{N(t)} \, \psi_{h,\sigma_t,j}(t,x)
\eeqn
with
\eqn
    \psi_{h,\sigma_t,j}(t,x)&:=&e^{itH(\sigma_t)}\int_{\Gamma_j(t)} dp \;
     W_{ \sigma_t}(V_j,t)e^{-i(p-P_f)x} \,
    \\
    &&\hspace{2.5cm}
    e^{i\widetilde\gamma_{\sigma_t}(V_j,\nabla_pE(p,\sigma_t),t)}
    \, h(p)  \, e^{-itE(p,\sigma_t)} \, \Psivec(p,\sigma_t) \;,
    \nonumber
\eeqn
and
\eqn
    V_j \; := \; \nabla_pE(p_j,\sigma) \;,
\eeqn
where $p_j$ is the center of the cell $\Gamma_j(t)$.
Here, $W_{ \sigma_t}(V_j,t)$ is defined as the operator obtained after
replacing $\nabla_pE(p ,\sigma)$ by $V_j$ in $W_{p,\sigma_t}(t)$, and
$\widetilde\gamma_{\sigma_t}(V_j,\nabla_pE(p,\sigma_t),t)$ is a
scalar phase factor.

The main result of  \cite{pi2}, formulated for the model of non-relativistic QED
studied here, can be stated as follows.

Let
\eqn
    \sigma_t \; \sim \; t^{-\beta} \;,
    \label{eq:sig-t-def-1}
\eeqn
for $\beta>1$ sufficiently large, and  $N(t)$, $T_n^{(\e)}$ as in (\ref{eq:Nt-def}), with
$\e$ sufficiently small.
Then the limit
\eqn
    \psi_h^{(out)} \; = \; s-\lim_{t\rightarrow\infty} \psi_{h,\sigma_t}(t)
    \label{eq:scatt-lim}
\eeqn
exists in the one-particle Hilbert space $\H=\H_{el}\otimes\Fo$,
for the model of non-relativistic QED defined in Section {\ref{sec:model-def}}.
A similar result holds for $t\rightarrow-\infty$, yielding a state $\psi_h^{(in)}$.

The vectors $\psi_h^{(in/out)}$ are {\em infraparticle scattering states}.

The proof strategy of \cite{pi2} comprises two main steps, which can
be sketched as follows.

\subsection{Step 1: Control of the norm}

This step consists in proving that the norm
$\|\psi_{h,\sigma_t}(t)\|_{\H}$ is uniformly bounded in $t$, and that, in fact,
\eqn
        \lim_{t\rightarrow\infty}\Big\| \, \psi_{h,\sigma_t}(t) \, \Big\|_{\H} \; =\; \|h\|_{L^2} \;.
        \label{eq:normlim-1}
\eeqn
Introducing the matrix elements
\eqn
        M_{i,j}(t) \; := \; \Bra \psi_{h,\sigma_t,i}(t) \,,\, \psi_{h,\sigma_t,j}(t)\Ket_{\H} \;,
\eeqn
one easily sees that the sum over diagonal terms, $i=j$, yields the right hand
side of (\ref{eq:normlim-1}), in the limit $t\rightarrow\infty$.
The off-diagonal matrix elements are shown to satisfy
\eqn
    | \, M_{i,j}(t) \, | \; < \; c(t) \; \; \; \; , \; \; i\neq j \;,
    \label{eq:offdiag-est-1}
\eeqn
where
\eqn
    c(t) \, N(t)^2 \; \searrow \; 0 \;,
\eeqn
as $t\rightarrow\infty$, so that
\eqn
    \lim_{t\rightarrow\infty}\sum_{i\neq j} \, | \, M_{i,j}(t) \, | \; = \; 0 \;.
\eeqn
Since the centers of the cells $\Gamma_j(t)$ label distinct
asymptotic velocities of infraparticle states,
this result implies that, asymptotically, the latter become mutually orthogonal, for $i\neq j$.
One uses here dispersive estimates for the free infraparticle propagation, which are derived from
the uniform bounds on $\partial_p^2E(p,\sigma)$, for $\sigma\geq0$.
For further details, see \cite{pi2}.

\subsection{Step 2: Strong convergence}

In this step, one proves that $\{\psi_{h,\sigma_{t}}(t)\}_t$ defines a Cauchy sequence in the one-particle
Hilbert space $\H$.
To this end, let $t_2>t_1\gg1$. The main result of \cite{pi2} is an estimate of the form
\begin{align}
    \Big\| \, \psi_{h,\sigma_{t_2}}(t_2) \, - \, \psi_{h,\sigma_{t_1}}(t_1) \, \Big\|_{\H} \; < \; t_1^{-\delta}
    \; \; , \; \; \delta>0 \;.
    \label{eq:cauchy-diff-1}
\end{align}
The proof in \cite{ pi2} is organized as follows.

Let $\psi_{h,\sigma_t,\Gamma(t')}(s)$ denote the vector obtained from
$\psi_{h,\sigma_{t'}}(t')$ by first replacing $\sigma_{t'}\rightarrow\sigma_t$ and then $t'\rightarrow s$,
while keeping  the cell decomposition
\eqn
    \Gamma(t') \; := \; \{ \, \Gamma_{j}(t') \, \}_{j =1}^{N(t')}
\eeqn
corresponding to time $t'$ fixed.

Assuming $t_2>t_1\gg1$, the left hand side of (\ref{eq:cauchy-diff-1}) is estimated by
\begin{align}
    \Big\| \, \psi_{h,\sigma_{t_2},\Gamma(t_2)}(t_2) \, - \,
    \psi_{h,\sigma_{t_1},\Gamma(t_1)}(t_1) \, \Big\|_{\H}
    \; \leq \; (I) \, + \, (II) \, + \, (III)
\end{align}
with the following definitions.
\begin{itemize}
\item
The term
\eqn
    (I) \; := \; \Big\| \, \psi_{h,\sigma_{t_2},\Gamma(t_2)}(t_2)
    \, - \, \psi_{h,\sigma_{t_2},\Gamma(t_1)}(t_2) \, \Big\|_{\H}
\eeqn
is the error made by replacing $\Gamma(t_2)$ by the coarser
cell decomposition $\Gamma(t_1)$ in
$\psi_{h,\sigma_{t_2},\Gamma(t_2)}(t_2)$, while keeping the infrared cutoff and the argument $t_2$ fixed.
One can control $(I)$ similarly as the off-diagonal terms in
(\ref{eq:offdiag-est-1}).
\item
The term
\eqn
    (II) \; := \; \Big\| \, \psi_{h,\sigma_{t_2},\Gamma(t_1)}(t_2)
    \, - \, \psi_{h,\sigma_{t_1},\Gamma(t_1)}(t_2) \, \Big\|_{\H}
\eeqn
is the error made by subsequently changing the infrared cutoff from $\sigma_{t_2}$ to $\sigma_{t_1}$
in $\psi_{h,\sigma_{t_2},\Gamma(t_1)}(t_2)$.
It admits a bound that
involves a positive power of $\sigma_{t_1}=t_1^{-\beta}$.
\item
The term
\eqn
    (III) \; := \; \Big\| \, \psi_{h,\sigma_{t_1},\Gamma(t_1)}(t_2)
    \, - \, \psi_{h,\sigma_{t_1},\Gamma(t_1)}(t_1) \, \Big\|_{\H}
\eeqn
is the left hand side of (\ref{eq:cauchy-diff-1}) with $\psi_{h,\sigma_{t_2},\Gamma(t_2)}(t_2)$
replaced by $\psi_{h,\sigma_{t_1},\Gamma(t_1)}(t_2)$.
To bound $(III)$, one applies Cook's argument to
\eqn
    \psi_{h,\sigma_{t_1},j_1}(t_2) \, - \,
    \psi_{h,\sigma_{t_1},j_1}(t_1) \; = \; \int_{t_1}^{t_2} ds \, \partial_s \,
    \psi_{h,\sigma_{t_1},j_1}(s) \;.
    \label{eq:psi-Duhamel-1}
\eeqn
This is the most involved part of the analysis,
and the integrand on the right hand side of (\ref{eq:psi-Duhamel-1}) must be subdivided
into many different terms for which one can either prove rapid decay in $s$ or
(asymptotically) precise cancellations.
\end{itemize}




\subsection{Modifications of \cite{pi2} for QED}

Most of the constructions in \cite{pi2} can be adopted
directly to yield the corresponding ones in non-relativistic QED.
The following minor modifications are necessary.

\begin{itemize}
\item
The infrared regularization is implemented by a sharp
cutoff $\chi(|k|>\sigma)$ in  \cite{pi2}.
It must be replaced by an infrared regularization $\cuts(|k|)$ which is zero at $|k|=0$,
but non-zero in an open
neighborhood of $|k|=0$.
Implementing this modification in \cite{pi2} (invoking results of \cite{ch1,ch2},
instead of \cite{fr, pi}) is straightforward.

\item
The dressing transformations in \cite{pi2} are slightly different from the
ones used in non-relativistic QED. In \cite{pi2},
the integral kernel corresponding to $v_{p,\sigma,\lambda}(k)$ has the form
\eqn
        \g \, \frac{\chi(\sigma<|k|<1)}{|k|^{\frac12}} \, \frac{1}{|k|-k\cdot\nabla_pE(p,\sigma)} \;,
\eeqn
while, here, there is an additional factor $\nabla_pE(p,\sigma)\cdot\e_\lambda(k)$;
see (\ref{eq:vpkernel-def}).
This does not lead to any non-trivial changes of the considerations in \cite{pi2}.
\end{itemize}

However, some other modifications are less straightforward,
due to the more complicated interaction term of non-relativistic QED.

\begin{itemize}
\item
In the application of Cook's method, there is a derivative
\eqn
        \partial_s\Big(e^{isH(\sigma_t)}W_{\sigma_t}(V_j,s)e^{-isH(\sigma_t)}\Big)
\eeqn
which contains a term of the form
\eqn
        i e^{isH(\sigma_t)} [H(\sigma_t)-H_f,W_{\sigma_t}(V_j,s)]e^{-isH(\sigma_t)}
\eeqn
(we recall that the interaction term in $H(\sigma)$ depends on $x$).
Due to the linear coupling in Nelson's model, the above commutator is given by
\eqn
        [H(\sigma_t)-H_f,W_{\sigma_t}(V_j,s)]=W_{\sigma_t}(V_j,s)\phi_{\sigma_t,V_j}(x,s) \;,
\eeqn
where $\phi_{\sigma_t,V_j}(x,s)$
is a scalar function that has rapid decay in $s$.

For QED, $\phi_{\sigma_t,V_j}(x,s)$ is replaced by
an operator linear in $\nabla_p H(p,\sigma_t)$ (for total momentum $p$).
The modifications arising here are technically somewhat demanding  and involve an
application of the uniform bounds on the renormalized electron mass.
\end{itemize}

A more detailed analysis of scattering theory along the lines of \cite{pi2}
would be appropriate.

\section{Proofs of Theorems {\ref{thm:main-1}} and {\ref{thm:main-2}}}

Our proofs follow closely \cite{fr}, where the statements of Theorems
{\ref{thm:main-1}} and {\ref{thm:main-2}} were established for Nelson's model.

In our proofs of Theorem {\ref{thm:main-1}} and part 2 of Theorem {\ref{thm:main-2}},
the first step is to employ the usual "pull-through formula",
which yields an explicit expression for
$a_\lambda(k)\Psivec(p,\sigma)$ in terms of $\Psivec(p,\sigma)$.
However, this is not the end of the story,
in contrast to \cite{fr}, where the result corresponding to Theorem
{\ref{thm:main-2}} for Nelson's model was established.
In non-relativistic QED,
the different coupling structure in the Hamiltonian $H(p,\sigma)$
poses considerable difficulties. Our
method involves application of the uniform bounds (\ref{eq:renmass-bd})
on the renormalized electron mass, which has only recently become available.

Our main technical result is formulated in the following
proposition.

\begin{proposition}
\label{prop:aPsi-main-error}
Under the hypotheses of Theorem {\ref{thm:main-recall}},
the vector $a_\lambda(k)\Psivec(p,\sigma)$ can be decomposed into
\eqn
    a_\lambda(k)\Psivec(p,\sigma) \; = \; \Phi_1(p,\sigma;k,\lambda)
    \, + \, \Phi_2(p,\sigma;k,\lambda)\;,
\eeqn
where
\begin{align}
    \Phi_1(p,\sigma;k,\lambda) \; = \;
    - \, \g \, \e_\lambda(k)\cdot\nabla_p E(p,\sigma) \,
    \frac{\cuts(|k|)}{|k|^{\frac12}}\frac{1}{|k|-k\cdot\nabla_p E(p,\sigma)} \,
    \Psivec(p,\sigma)
\end{align}
and
\eqn
    \|\Phi_2(p,\sigma;k,\lambda)\| \; \leq \; c \, \g \, \frac{\cuts(|k|)}{|k| }   \;,
\eeqn
for a constant $c$ that is independent of $\sigma$ and $\gs$.
\end{proposition}

The uniform bound on the renormalized electron mass (\ref{eq:renmass-bd}) enters the
estimate for the vector $\Phi_2(p,\sigma;k)$ . (We recall that $\cuts$ denotes the
cutoff function in (\ref{eq:A-B-def-1}).)

\subsection{Proof of Theorem {\ref{thm:main-1}}}

The statement of Theorem {\ref{thm:main-1}} is an immediate consequence of
Proposition {\ref{prop:aPsi-main-error}}.

\subsection{Proof of Theorem {\ref{thm:main-2}}, Part 1}

For the existence of a convergent subsequence, we refer to \cite{fr}.
The proof comprises the following main steps.

Let $K_{\rho}:=\{k\in\R^3\,\big|\, |k|\,\geq\,\rho\}$
for $0<\rho< 1$, and let $\Fo_{\rho }$ denote the Fock space over the
one-photon Hilbert space $L^2(K_{\rho })\otimes\C^2$. Let $\alg_{\rho }$ denote the $C^*$-algebra
of bounded operators on $\Fo_{\rho }$.

One first establishes the existence of an operator $C_\rho$ affiliated with $\alg_{\rho }$
which has a {\em compact resolvent} on $\Fo_{\rho }$, and which satisfies
\eqn
        \omega_{p,\sigma}(C_{\rho}) \; < \; M(\rho) \; < \; \infty
\eeqn
{\em uniformly} in $\sigma>0$.  For instance, the operator
\eqn
        C_{\rho } \; := \; \sum_{\lambda=+,-} \, \int_{ |k|\geq\rho}
        dk\, a^*_\lambda(k) \,
        \big[ \, -   \Delta_k\,+\, |k|^2 \, \big] \,  a_\lambda(k)
\eeqn
has these properties in the present case
(see also \cite{fr,fr2} and \cite{glja}).

It follows that $\{\omega_{p,\sigma}\Big|_{\alg_{\rho}}\}_{\sigma>0}\subset\alg_{\rho}^*$
is norm compact, see \cite{glja}. The dual $\alg_{\rho}^*$
of $\alg_{\rho}$ is a Banach space, because
$\alg_{\rho}$ is a von Neumann algebra.
Hence, for any sequence $\{\sigma_j\}_{j=0}^\infty$ converging to zero,
there exists a subsequence $\{\sigma_{j_l}\}_{l=0}^\infty$ converging to zero
such that $\{\omega_{p,\sigma_{j_l}}\}_{l=0}^\infty$ converges to a normal state
$\omega_{p}^{(\rho)}$ on $\alg_{\rho}$.

Choosing $\rho_n=\frac{1}{n}$ for $n\in\N$, we get, by Cantor's diagonal
procedure, a subsequence $\{\sigma_{j_l}\}_{l=0}^\infty$ converging to 0 such that
$\{\omega_{p,\sigma_{j_l}}\}_{l=0}^\infty$ converges on $\alg_{\frac1n}$, for all
$n<\infty$. Hence, $\{\sigma_{p,\sigma_{j_l}}\}_{l=0}^\infty$ converges on
$\bigvee_{n}\alg_{\frac1n}$, and thus on $\alg$, to a state $\omega_p$ on $\alg^*$.
$\omega_{p}^{(\rho)}=\omega_p\big|_{\alg_{\rho}}$ is a normal state.

\subsection{Proof of Theorem {\ref{thm:main-2}}, part 2.}
This is an immediate consequence of Proposition {\ref{prop:aPsi-main-error}}.

Indeed, we have that
\eqn
    \Big|\bra\Psivec(p,\sigma)\,,\,\Phi_1(p,\sigma;k,\lambda)\ket\Big|^2
    &=&
    \bra\Phi_1(p,\sigma;k,\lambda)\,,\,\Phi_1(p,\sigma;k,\lambda)\ket  \;,
\eeqn
since $\Phi_1(p,\sigma;k,\lambda)$ is a scalar multiple of $\Psivec(p,\sigma)$, and $\|\Psivec(p,\sigma)\|=1$. Therefore,
\eqn
    \lefteqn{
    \bra\Psivec(p,\sigma)\,,\,a^*_\lambda(k)\,a_\lambda(k)\Psivec(p,\sigma)\ket
    }
    \nonumber\\
    &=&\bra\Phi_1(p,\sigma;k,\lambda)\,,\,\Phi_1(p,\sigma;k,\lambda)\ket + \rho_1(p,\sigma;k,\lambda)
    \nonumber\\
    &=&\Big|\bra\Psivec(p,\sigma)\,,\,\Phi_1(p,\sigma;k,\lambda)\ket\Big|^2 + \rho_1(p,\sigma;k,\lambda)
    \nonumber\\
    &=&\Big|\bra\Psivec(p,\sigma)\,,\,a_\lambda(k)\Psivec(p,\sigma)\ket\Big|^2 + \rho_1(p,\sigma;k,\lambda)
    -\rho_2(p,\sigma;k,\lambda)
\eeqn
where
\eqn
    \rho_1(p,\sigma;k,\lambda)&=&\bra\Phi_1(p,\sigma;k,\lambda)\,,\,\Phi_2(p,\sigma;k,\lambda)\ket
    +\bra\Phi_2(p,\sigma;k,\lambda)\,,\,\Phi_1(p,\sigma;k,\lambda)\ket
    \nonumber\\
    &+&\bra\Phi_2(p,\sigma;k,\lambda)\,,\,\Phi_2(p,\sigma;k,\lambda)\ket
\eeqn
and
\eqn
    \rho_2(p,\sigma;k,\lambda)&=&\bra\Psivec(p,\sigma)\,,\,\Phi_1(p,\sigma;k,\lambda)\ket
    \bra\Phi_2(p,\sigma;k,\lambda)\,,\,\Psivec(p,\sigma)\ket
    \nonumber\\
    &+&\bra\Psivec(p,\sigma)\,,\,\Phi_2(p,\sigma;k,\lambda)\ket
    \bra\Phi_1(p,\sigma;k,\lambda)\,,\,\Psivec(p,\sigma)\ket
    \nonumber\\
    &+&\Big|\bra\Psivec(p,\sigma)\,,\,\Phi_2(p,\sigma;k,\lambda)\ket\Big|^2 \;.
\eeqn
Clearly,
\begin{align}
    |\rho_1(p,\sigma;k,\lambda)|\,,\,|\rho_2(p,\sigma;k,\lambda)|&\leq \;
    2\|\Phi_1(p,\sigma;k,\lambda)\|\,\|\Phi_2(p,\sigma;k,\lambda)\|
    \nonumber\\
    &\hspace{1cm}+\|\Phi_2(p,\sigma;k,\lambda)\|^2
    \nonumber\\
    &\leq\;c\gs  |\nabla_p E(p,\sigma)| \frac{\cuts^2(|k|)}{|k|^{\frac52} }
    \, + \, c'\gs   \frac{\cuts^2(|k|)}{|k|^{2} }  \;.
\end{align}
This proves the claim.

\subsection{Proof of Theorem {\ref{thm:main-2}}, part 3.} We sketch the
proof, and refer to Lemma 3.1 in \cite{fr} for details (see also \cite{brro1,dedoru,eckfr}).

We consider the {\em  coherent *-automorphisms}
\eqn
    \alpha_{p,\sigma}(A) =  W_{p,\sigma} A W_{p,\sigma}^*  \; \; , \; \; A\in\alg \;,
\eeqn
where
\eqn
        W_{p,\sigma} \; = \;  \exp\Big[ i\sum_{\lambda=+,-}
        \Pi_\lambda( v_{p,\sigma,\lambda})\Big]   \;,
\eeqn
see (\ref{eq:coh-state-1}),
and $\Pi_\lambda(f)=i(a_\lambda(f)-a_\lambda^*(f))$.
In the limit $\sigma\searrow0$, the states
\eqn
    \mu_{p,\sigma} :=  \omega_p(\alpha_{p,\sigma}( \, \cdot \, ))
\eeqn
converge to
\eqn
    \mu_p  = \omega_p(\alpha_p(\,\cdot\,)) \;,
\eeqn
where $\alpha_{p}(A)  = n-\lim_{\sigma\searrow0} \alpha_{p,\sigma} (A)$,
for $A\in\alg$; see \cite{fr}.

Next, one proves that the representation $\pi_{\mu_p}=\pi_p\circ \alpha_p$
admits a {\em positive, selfadjoint} number operator.
This implies that $\pi_{\mu_p}$ is
quasi-equivalent to the Fock representation, for $0\leq|p|<\puppbd$, \cite{dedoru}.
To this end, we define  the local number operators
\eqn
    N_{\rho } \; := \; \sum_{\lambda=+,-}\int_{ |k|>\rho } dk \, a_\lambda^*(k) \, a_\lambda(k)
    \;\;\;,\;\;\;\;{\rm for}\;\rho>0 \;,
    \label{eq:loc-number-op}
\eeqn
where $\exp[itN_\rho]\in\alg_{\rho}$.
Let $\H_{\mu_p}$ denote the Hilbert space and $\Omega_{\mu_p}\in\H_{\mu_p}$ the cyclic vector
corresponding to $\mu_p$ by GNS construction.

One can show that $\pi_{\mu_p}(\exp[itN_\rho])\pi_{\mu_p}(A)\Omega_{\mu_p}$
converges strongly, as $\rho\searrow0$, for all $A\in\bigvee_{\rho>0}\alg_\rho$, and
all $t\in\R$. The limit of  $\pi_{\mu_p}(\exp[itN_\rho])$, as $\rho\searrow0$,
$t\in\R$, defines a strongly continuous unitary group on $\H_{\mu_p}$.
Its generator defines a positive, selfadjoint number operator on $\H_{\mu_p}$.

Since $A\in\bigvee_{\rho>0}\alg_\rho$, there is some $\widetilde\rho>0$ such that
$A\in\alg_{\widetilde \rho}$. Let $\rho'\leq\rho\leq\widetilde \rho$, and let
\eqn
    N_{\rho',\rho }:= \sum_{\lambda=+,-}\int_{ \rho'\leq|k|\leq\rho } dk \, a_\lambda^*(k) \, a_\lambda(k)\;.
\eeqn
Then,
\eqn
    &&\|\pi_{\mu_p}((e^{itN_\rho}-e^{itN_{\rho'}})A)\Omega_{\mu_p}\|^2
    \nonumber\\
    &&\hspace{2cm}=\;
    2\mu_p(A^*A)-\mu_p(A^*e^{itN_{\rho',\rho}}A)-\mu_p(A^*e^{-itN_{\rho',\rho}}A)\;.
\eeqn
Using that
\eqn
    \pi_{\mu_p}(\1-e^{itN_{\rho',\rho}})=-i\int_0^t\pi_{\mu_p}(e^{isN_{\rho',\rho}}N_{\rho',\rho}) \;,
\eeqn
a straightforward calculation shows that
\eqn
    &&|\mu_p(A^*A)-\mu_p(A^*e^{itN_{\rho',\rho}}A)|
    \nonumber\\
    &&\hspace{2cm}\leq \;
    |t|\,\|A\|^2\sum_{\lambda}\int_{\rho'\leq|k|\leq\rho}dk\Big[\omega_p(a^*_\lambda(k)a_\lambda(k))
    -|\omega_p(a_\lambda(k))|^2\Big]
    \nonumber\\
    &&\hspace{2cm}\leq\;|t|\,\|A\|^2\,|\rho-\rho'|^{\frac12} \;,
\eeqn
which tends to zero as $\rho\searrow0$. In the last step, we used (\ref{eq:state-int-1}).

Our results imply that $\pi_p$ is quasi-equivalent to the coherent representation corresponding to
(\ref{eq:coh-state-1}) by the GNS construction.

\subsection{Proof of Theorem {\ref{thm:main-2}}, part 4.}

Theorem {\ref{thm:main-recall}}
implies  that  $\nabla_p E(p,\sigma)\neq0$ if and only if $p\neq 0$, for $|p|<\puppbd$.
Thus if $p\neq0$, (\ref{eq:Nf-divergence-sharp}) in Theorem {\ref{thm:main-1}} implies that
\eqn
    \omega_{p,\sigma}(N_f)
    &=&\bra\Psivec(p,\sigma)\,,\,N_f\Psivec(p,\sigma) \ket
    \nonumber\\
    &\geq&c \gs( 1 + |\nabla_pE(p,\sigma)|^2\log\frac1\sigma )
\eeqn
which diverges to $\infty$ as $\sigma\searrow0$.
Hence, (\ref{eq:num-lim-1}) follows. However, if $p=0$,
one gets (\ref{eq:num-lim-2}) from $\nabla_p E(p,\sigma)=0$.

The local Fock properties of $\omega_p$ are derived from the following considerations.

Using Proposition {\ref{prop:aPsi-main-error}}, it is easy to see that
\eqn
    \omega_{p,\sigma}(N_{\rho})&\leq&
    2\sum_\lambda \int_{ |k|>\rho } dk
    \Big[\|\Phi_1(p,\sigma;k,\lambda)\|^2
    +\|\Phi_2(p,\sigma;k,\lambda)\|^2\Big] \,
    \nonumber\\
    &<&C(\rho) \;,
\eeqn
uniformly in $\sigma\geq0$. By similar considerations as in the proof of
part 3 of  Theorem {\ref{thm:main-2}}, one concludes
that the representation of $\alg_\rho$ corresponding to $\omega_p$ by GNS construction
is quasi-equivalent to the Fock representation for every $\rho>0$.

Let $B\subset\R^3$ denote a bounded region in physical $x$-space, and
let $\alg(B)$ denote the corresponding local algebra. Then, the restriction
of $\omega_p$ to $\alg(B)$ defines a GNS representation of $\alg(B)$
which is quasi-equivalent to the Fock representation.
This can be shown by a straightforward adaptation of
results in \cite{glja} to the present model.

\section{Proof of Proposition {\ref{prop:aPsi-main-error}}}

It remains to prove Proposition {\ref{prop:aPsi-main-error}}, our key
analytical result in this paper. To this end, we
first derive the following representation of $a_\lambda(k)\Psivec(p,\sigma)$.

\begin{lemma}
Assume that $0<|k|<1$, $|p|< \puppbd$, and $\gs<\gs_0$
(where $\gs_0$ denotes the same constant as in Theorem {\ref{thm:main-recall}}).
Let $E(p,\sigma)$ denote the ground state eigenvalue of $H(p,\sigma)$, and let
$\Psivec(p,\sigma)$ be an eigenvector in the corresponding two-dimensional eigenspace.

Then the operator $H(p-k)+|k|-E(p,\sigma)$ is invertible, and
\eqn
    a_\lambda(k)  \Psivec(p,\sigma)
    \;=&-&\frac{1}{ H(p-k,\sigma) +|k|-E(p,\sigma) }
    \nonumber\\
    &&\Big[ \g  \frac{\cuts(|k|)}{|k|^{\frac12}}\e_\lambda(k)\cdot\nabla_p H(p,\sigma)
    \label{alk-psi-GLL-1}\\
    &&\hspace{2cm}+\;
    \g \frac{\cuts(|k|)}{|k|^{\frac12}} \tau\cdot (k\wedge \e_\lambda(k))
    \Big] \Psivec(p,\sigma) \;.
    \nonumber
\eeqn
In the scalar case, $E(p,\sigma)$ is a simple eigenvalue, and the magnetic term
(proportional to $\tau$) is absent.

Moreover, the a priori bound
\eqn
    \|a_\lambda(k)  \Psivec(p,\sigma)\| \; \leq \; c \g \frac{\cuts(|k|)}{|k|^{\frac32}}
    \Big[\sqrt{ p^2 +c'\alpha}+ |k|\Big]
    \label{eq:apriori-bd}
\eeqn
holds, where the constants $c$ and $c'$ are independent of $\gs$ and $\sigma$.
\end{lemma}

\begin{proof}
We recall the definition of the fiber Hamiltonian
\eqn
    H(p,\sigma) \; = \; \frac12 (p-P_f-\g A_\sigma)^2 + \g \tau\cdot B_\sigma + H_f \;.
\eeqn
The "pull-through formula" says that
\eqn
    a_\lambda(k)H(p,\sigma)
    &=&
    \Big( \frac12 (p-P_f-k-\g A_\sigma)^2
    +
    \g \tau\cdot B_\sigma + H_f +|k| \Big) a_\lambda(k)
    \nonumber\\
    &&\hspace{2cm}-
    \g \frac{\cuts(|k|)}{|k|^{\frac12}} \, \e_\lambda(k)\cdot(p-P_f-\g A_\sigma)
    \nonumber\\
    &&\hspace{2cm}+
    \g \frac{\cuts(|k|)}{|k|^{\frac12}} \, \tau\cdot (ik\wedge \e_\lambda(k)) \;,
\eeqn
where $\tau=(\tau_1,\tau_2,\tau_3)$ is the vector of Pauli matrices. We observe that
\eqn
    \nabla_p H(p,\sigma) \; = \;  p \, - \, P_f \, - \, \g A_\sigma \;,
\eeqn
and that
\eqn
    \e_\lambda(k)\cdot \nabla_p H(p-k,\sigma) \; = \; \e_\lambda(k)\cdot \nabla_p H(p,\sigma)
\eeqn
since $\e_\lambda(k)\cdot k=0$, by the Coulomb gauge condition.
Hence
\eqn
    a_\lambda(k) E(p,\sigma) \Psivec(p,\sigma)
    &=&
    a_\lambda(k) H(p,\sigma)  \Psivec(p,\sigma)
    \nonumber\\
    &=&
    \Big[ \big( H(p-k,\sigma) +|k|\big) a_\lambda(k)
    \nonumber\\
    &&\hspace{2cm}+
    \g  \frac{\cuts(|k|)}{|k|^{\frac12}} \, \e_\lambda(k)\cdot\nabla_p H(p,\sigma)
    \label{pull-through-1}\\
    &&\hspace{2cm}+
    \g \frac{\cuts(|k|)}{|k|^{\frac12}} \, \tau\cdot (ik\wedge \e_\lambda(k))
    \Big] \Psivec(p,\sigma)\;,
    \nonumber
\eeqn
so that
\eqn
    \lefteqn{
    \Big[ H(p-k,\sigma) +|k|-E(p,\sigma)\Big]a_\lambda(k)  \Psivec(p,\sigma)
    }
    \label{eq:aux-pullthrough-ev}\\
    &=&
    -
    \Big[ \g  \frac{\cuts(|k|)}{|k|^{\frac12}} \, \e_\lambda(k)\cdot\nabla_p H(p,\sigma)
    \, + \,
    \g \frac{\cuts(|k|)}{|k|^{\frac12}} \, \tau\cdot (ik\wedge \e_\lambda(k))
    \Big] \Psivec(p,\sigma) \;.
    \nonumber
\eeqn
Furthermore, the bounds
\eqn
    H(p-k,\sigma)+|k|-E(p,\sigma) \; \geq \; E(p-k,\sigma)+|k|-E(p,\sigma)
    \; > \;  \frac{|k|}{10}
    \label{eq:Hshift-lowerbd-2}
\eeqn
follow from (\ref{eq:Hshift-lowerbd-1}) below.
Hence, $H(p-k,\sigma)+|k|-E(p,\sigma)$
(see left side of (\ref{eq:aux-pullthrough-ev})) is invertible,
for any $0<|k|<1$ and $|p|< \puppbd$.

We conclude that
\eqn
    a_\lambda(k)  \Psivec(p,\sigma)
    \;=&-&\frac{1}{ H(p-k,\sigma) +|k|-E(p,\sigma) }
    \nonumber\\
    &&\Big[ \g \, \frac{\cuts(|k|)}{|k|^{\frac12}}\e_\lambda(k)\cdot\nabla_p H(p,\sigma)
    \\
    &&\hspace{2cm}+\;
    \g \, \frac{\cuts(|k|)}{|k|^{\frac12}} \tau\cdot (ik\wedge \e_\lambda(k))
    \Big] \Psivec(p,\sigma) \;,
    \nonumber
\eeqn
as claimed.

Moreover, (\ref{eq:Hshift-lowerbd-2}) immediately implies the a priori bound (\ref{eq:apriori-bd}).
\end{proof}

Proof of Proposition {\ref{prop:aPsi-main-error}}.

\begin{proof}
We note that
\eqn
    (\nabla_p H)(p,\sigma)  \Psivec(p,\sigma) &=& \nabla_p (H(p,\sigma)  \Psivec(p,\sigma))
    \, - \,  H(p,\sigma)  \nabla_p \Psivec(p,\sigma)
    \nonumber\\
    &=& \nabla_p (E(p,\sigma)  \Psivec(p,\sigma))
    \, - \,  H(p,\sigma)  \nabla_p \Psivec(p,\sigma)
    \nonumber\\
    &=& (\nabla_p E)(p,\sigma) \Psivec(p,\sigma)
    \\
    &&\hspace{2cm}- \,  (H(p,\sigma) - E(p,\sigma)) \nabla_p \Psivec(p,\sigma) \;.
    \nonumber
\eeqn
From (\ref{alk-psi-GLL-1}), we get
\eqn
    a_\lambda(k)  \Psivec(p,\sigma)
    &=&(I)+(II) \;,
\eeqn
where
\eqn
    (I)&=&-\g  \frac{\cuts(|k|)}{|k|^{\frac12}}(\e_\lambda(k)\cdot\nabla_p E(p,\sigma))
    \frac{1}{ H(p-k,\sigma) +|k|-E(p,\sigma) }   \Psivec(p,\sigma)
    \nonumber\\
    (II)&=&
    \g  \frac{\cuts(|k|)}{|k|^{\frac12}}\frac{1}{ H(p-k,\sigma) +|k|-E(p,\sigma) }
    \nonumber\\
    &&\hspace{2cm}
    \Big[(H(p,\sigma)-E(p,\sigma))  \e_\lambda(k)\cdot\nabla_p \Psivec(p,\sigma)
    \nonumber\\
    &&\hspace{4cm}
    -\; \tau\cdot (ik\wedge \e_\lambda(k))\Psivec(p,\sigma) \Big]  \;.
\eeqn
Let us first bound $(II)$.

To this end, we first prove that for $0<|k|<1$ and $|p|<\puppbd$,
\eqn
    \Big\|(H(p,\sigma)-E(p,\sigma))\frac{1}{H(p-k,\sigma)+|k|-E(p,\sigma)}\Big\|_{op}\leq 3  \;.
    \label{eq:resolv-relat-bds}
\eeqn
We note that
\eqn
    H(p-k,\sigma) &=& H(p,\sigma) +\frac{k^2}{2} - k\cdot \nabla_p H(p,\sigma)
    \label{eq:Hshift-decomp}
\eeqn
so that
\eqn
    H(p-k,\sigma)-E(p,\sigma) &=& H(p,\sigma)-E(p,\sigma) - k\cdot\nabla_p H(p,\sigma)+\frac{k^2}{2}
    \nonumber\\
    &\geq&H(p,\sigma)-E(p,\sigma) -  \frac{k^2}{2\delta} -  \frac{\delta}{2}
    (\nabla_p H(p,\sigma))^2 +\frac{k^2}{2}
    \nonumber\\
    &\geq&(1-\frac23 |k|)(H(p,\sigma)-E(p,\sigma))+\frac{k^2}{2}-\frac{3}{4}|k|
    \\
    &&\hspace{2cm}
    +\frac23|k|(H_f+\g\tau\cdot B_\sigma-E(p,\sigma)) \;,
    \nonumber
\eeqn
using the Schwarz inequality with $\delta=\frac23|k|\leq\frac23$.
From
\eqn
    | B_\sigma | \leq c \sqrt{1+H_f}\;,
\eeqn
the operator on the last line is bounded by
\eqn
    &&\frac23 |k|\Big[\chi(H_f\geq1)(H_f-\g c \sqrt{1+H_f})-c\g-E(p,\sigma)\Big]
    \nonumber\\
    &&\hspace{3cm}\geq\;
    - \frac23|k|\Big[\frac12\left(\puppbd\right)^2+c\gs\Big] \;,
\eeqn
using $E(p,\sigma)\leq\frac{p^2}{2}+c\gs$ for $|p|<\puppbd$.
Therefore
\eqn
    H(p-k,\sigma)-E(p,\sigma)+|k|
    &\geq&(1-\frac23|k|)(H(p,\sigma)-E(p,\sigma))+\frac{k^2}{2}
    \nonumber\\
    &&\hspace{2cm}+ |k|(1-\frac{3}{4}-\frac12\left(\puppbd\right)^2-c\gs)
    \nonumber\\
    &\geq&\frac13(H(p,\sigma)-E(p,\sigma))+\frac{|k|}{10} \;,
    \label{eq:Hshift-lowerbd-1}
\eeqn
for $|k|<1$. This implies (\ref{eq:resolv-relat-bds}).

It is then easy to see that
\eqn
    \|(II)\|&\leq&c \, \g \, \frac{\cuts(|k|)}{|k|^{\frac12}}\Big[
    \Big\|(H(p,\sigma)-E(p,\sigma))\frac{1}{H(p-k,\sigma)+|k|-E(p,\sigma)}\Big\|_{op}^{\frac12}
    \nonumber\\
    &&\hspace{3cm}
    \Big\|\frac{1}{H(p-k,\sigma)+|k|-E(p,\sigma)}\Big\|_{op}^{\frac12}
    \nonumber\\
    &&\hspace{4cm}
    \Big\|(H(p,\sigma)-E(p,\sigma))^{\frac12}\nabla_p \Psivec(p,\sigma))\Big\|
    \nonumber\\
    &&
    \hspace{2cm}+\Big\|\frac{1}{H(p-k,\sigma)+|k|-E(p,\sigma)}\Big\|_{op}|ik\wedge\e_\lambda(k)|\Big]
    \nonumber\\
    &\leq&c \, \g \, \frac{\cuts(|k|)}{|k|^{\frac12}}
    \Big[\frac{1}{|k|^{\frac12}} \, \Big| \, \frac{1}{m_{ren}(p,\sigma)}-1 \, \Big|^{\frac12}
    +1\Big]
    \nonumber\\
    &\leq&c \,\g\, \frac{\cuts(|k|)}{|k| }   \;,
\eeqn
where
\begin{align}
     m_{ren}(p,\sigma) &=&\Big[1-2\bra\nabla_p\Psivec(p,\sigma)\,,\,(H(p,\sigma)-E(p,\sigma))
    \nabla_p\Psivec(p,\sigma)\ket\Big]^{-1}
\end{align}
is the {\em renormalized electron mass}.

We recall from (\ref{eq:renmass-bd}) that
$|m_{ren}(p,\sigma)-1|<c\gs$, for $|p|<\puppbd$, uniformly in $\sigma\geq0$.

Next, we discuss the term $(I)$.
We use the resolvent identity and (\ref{eq:Hshift-decomp}) for
\eqn
    \frac{1}{H(p-k,\sigma)+|k|-E(p,\sigma)}&=&\frac{1}{H(p,\sigma)-E(p,\sigma)+|k|+\frac{k^2}{2}}
    \\
    &-&
    \frac{1}{H(p-k,\sigma)+|k|-E(p,\sigma)}  \;
    k\cdot\nabla_p H(p,\sigma)
    \nonumber\\
    &&\hspace{2cm}
    \frac{1}{H(p,\sigma)-E(p,\sigma)+|k|+\frac{k^2}{2}}\;.
    \nonumber
\eeqn
Accordingly,
\eqn
    (I)&=&(I_1)+(I_2) \;,
\eeqn
where
\eqn
    (I_1)&=&-\, \g \, \frac{\cuts(|k|)}{|k|^{\frac12}}(\e_\lambda(k)\cdot\nabla_p E(p,\sigma))
    \frac{1}{ H(p,\sigma) -E(p,\sigma) +|k|+\frac{k^2}{2}}   \Psivec(p,\sigma)
    \nonumber\\
    &=&- \, \g \, \frac{\cuts(|k|)}{|k|^{\frac12}}\frac{1}{|k|+\frac{k^2}{2}}
    (\e_\lambda(k)\cdot\nabla_p E(p,\sigma))\Psivec(p,\sigma)\;.
\eeqn
We note that the $L^2$-norm of this term {\em diverges logarithmically} in the limit $\sigma\searrow0$.

Moreover,
\eqn
    (I_2)&=&- \, \g \, \frac{\cuts(|k|)}{|k|^{\frac12}}(\e_\lambda(k)\cdot\nabla_p E(p,\sigma))
    \frac{1}{H(p-k,\sigma)+|k|-E(p,\sigma)} \;
    \nonumber\\
    &&\hspace{1cm}
    (k\cdot\nabla_p H(p,\sigma))\frac{1}{H(p,\sigma)-E(p,\sigma)+|k|+\frac{k^2}{2}}\Psivec(p,\sigma)
    \nonumber\\
    &=&- \, \g \, \frac{\cuts(|k|)}{|k|^{\frac12}}\frac{1}{|k|+\frac{k^2}{2}}
    (\e_\lambda(k)\cdot\nabla_p E(p,\sigma))
    \nonumber\\
    &&\hspace{1cm}
    \frac{1}{H(p-k,\sigma)+|k|-E(p,\sigma)} (k\cdot\nabla_p H(p,\sigma))   \Psivec(p,\sigma)
    \nonumber\\
    &=&(I_{21})+(I_{22})
\eeqn
with
\eqn
    (I_{21})&=&- \, \g \, \frac{\cuts(|k|)}{|k|^{\frac12}}\frac{1}{|k|+\frac{k^2}{2}}
    (\e_\lambda(k)\cdot\nabla_p E(p,\sigma))
    \nonumber\\
    &&\hspace{1cm}
    \frac{1}{H(p-k,\sigma)+|k|-E(p,\sigma)}(k\cdot\nabla_p E(p,\sigma))   \Psivec(p,\sigma)
    \nonumber\\
    &=&\frac{k\cdot\nabla_p E(p,\sigma)}{|k|+\frac{k^2}{2}} \cdot(I)
    \label{eq:I21-id-1}
\eeqn
and
\eqn
    (I_{22})&=&\g \, \frac{\cuts(|k|)}{|k|^{\frac12}}\frac{1}{|k|+\frac{k^2}{2}}
    (\e_\lambda(k)\cdot\nabla_p E(p,\sigma))
    \\
    &&\hspace{1cm}
    \frac{1}{H(p-k,\sigma)+|k|-E(p,\sigma)} (H(p,\sigma)-E(p,\sigma))
    k\cdot\nabla_p \Psivec(p,\sigma) \;.
    \nonumber
\eeqn
We find that
\eqn
    \|(I_{22})\|&\leq&3\g \, \frac{\cuts(|k|)}{|k|^{\frac12}}\frac{1}{|k|+\frac{k^2}{2}}
    |k||\nabla_p E(p,\sigma)|
    \nonumber\\
    &&\hspace{1cm}
    \Big\|\frac{1}{H(p-k,\sigma)+|k|-E(p,\sigma)}\Big\|_{op}^{\frac12}
    \nonumber\\
    &&\hspace{2cm}
    \Big\|(H(p,\sigma)-E(p,\sigma))\frac{1}{H(p-k,\sigma)+|k|-E(p,\sigma)}\Big\|_{op}^{\frac12}
    \nonumber\\
    &&\hspace{4cm}
    \Big\|(H(p,\sigma)-E(p,\sigma))^{\frac12}\nabla_p \Psivec(p,\sigma))\Big\|
    \nonumber\\
    &\leq&c \, \g \, \frac{\cuts(|k|)}{|k| } |\nabla_p E(p,\sigma)|\,
    \Big| \, \frac{1}{m_{ren}(p,\sigma)}-1 \, \Big|^{\frac12}
    \nonumber\\
    &\leq&c \, \gs \, \frac{\cuts(|k|)}{|k| } |\nabla_p E(p,\sigma)| \;,
\eeqn
using (\ref{eq:resolv-relat-bds}).

Hence, solving for $(I)$ (recalling that (\ref{eq:I21-id-1}) is a multiple of $(I)$),
\eqn
    (I)=\Big[1-\frac{k\cdot\nabla_p E(p,\sigma)}{|k|+\frac{k^2}{2}} \Big]^{-1}
    \Big[(I_1)+(I_{22})\Big]\;,
\eeqn
where
\eqn
    | \, k\cdot \nabla_p E(p,\sigma) \, | \; < \; |k|\,|p|\,(1+c\gs)
    \; < \; \frac{|k|}{2} \;,
\eeqn
for $|p|<\puppbd$ and $\gs$ sufficiently small, see (\ref{eq:derpE-bd-1}).
Noting that
\eqn
    \Big\|\Big[1-\frac{k\cdot\nabla_p E(p,\sigma)}{|k|+\frac{k^2}{2}} \Big]^{-1}
    (I_1)&+&
    \g \,
    \frac{\cuts(|k|)}{|k|^{\frac12}} \,
    \frac{\e_\lambda(k)\cdot\nabla_p E(p,\sigma)}{|k|- k\cdot\nabla_p E(p,\sigma)} \,
    \Psivec(p,\sigma)\Big\|
    \nonumber\\
    &<&c \g \, \frac{\cuts(|k|)}{|k|^{\frac12}} \,  \|\,\Psivec(p,\sigma)\,\| \;,
\eeqn
we find that
\eqn
    a_\lambda(k)\Psivec(p,\sigma) \; = \; \Phi_1(p,\sigma;k,\lambda) \, + \, \Phi_2(p,\sigma;k,\lambda)\;,
\eeqn
where
\begin{align}
    \Phi_1(p,\sigma;k,\lambda) \; = \;
    - \, \g \, \e_\lambda(k)\cdot\nabla_p E(p,\sigma) \,
    \frac{\cuts(|k|)}{|k|^{\frac12}}
    \frac{1}{|k|- k\cdot\nabla_p E(p,\sigma)} \,
    \Psivec(p,\sigma)
\end{align}
and
\eqn
    \|\Phi_2(p,\sigma;k,\lambda)\|&\leq&c \, \g \, \frac{\cuts(|k|)}{|k| }   \;.
\eeqn
This establishes Proposition {\ref{prop:aPsi-main-error}}.
\end{proof}

\subsection*{Acknowledgements}

We are grateful to A. Pizzo for some useful discussions of the work in \cite{pi2}.
T.C. was supported by NSF Grant  DMS-0524909.

\parskip = 0 pt
\parindent = 0 pt

\end{document}